\documentclass[twocolumn,english,final,showpacs,pra,superscriptaddress]{revtex4}
\usepackage[T1]{fontenc}
\usepackage[latin9]{inputenc}
\usepackage{verbatim}
\usepackage{textcomp}
\usepackage{amsmath}
\usepackage{graphicx}
\usepackage{amssymb}

\makeatletter
\@ifundefined{textcolor}{}
{%
 \definecolor{BLACK}{gray}{0}
 \definecolor{WHITE}{gray}{1}
 \definecolor{RED}{rgb}{1,0,0}
 \definecolor{GREEN}{rgb}{0,1,0}
 \definecolor{BLUE}{rgb}{0,0,1}
 \definecolor{CYAN}{cmyk}{1,0,0,0}
 \definecolor{MAGENTA}{cmyk}{0,1,0,0}
 \definecolor{YELLOW}{cmyk}{0,0,1,0}
 }

\usepackage{textcomp}
\usepackage{babel}

\usepackage{amsfonts}

\providecommand{\U}[1]{\protect\rule{.1in}{.1in}}
\makeatletter
\providecommand{\LyX}{L\kern-.1667em\lower.25em\hbox{Y}\kern-.125emX\@}
\@ifundefined{textcolor}{}
{
\definecolor{BLACK}{gray}{0}
\definecolor{WHITE}{gray}{1}
\definecolor{RED}{rgb}{1,0,0}
\definecolor{GREEN}{rgb}{0,1,0}
\definecolor{BLUE}{rgb}{0,0,1}
\definecolor{CYAN}{cmyk}{1,0,0,0}
\definecolor{MAGENTA}{cmyk}{0,1,0,0}
\definecolor{YELLOW}{cmyk}{0,0,1,0}
}
\makeatother
\makeatother

\makeatother

\usepackage{babel}

\makeatother

\usepackage{babel}

\makeatother

\usepackage{babel}

\makeatother

\usepackage{babel}

\begin{document}

\title{Performance comparison of dynamical decoupling sequences for a qubit
in a rapidly fluctuating spin-bath}

\author{Gonzalo A. \'{A}lvarez}

\email{galvarez@e3.physik.uni-dortmund.de}

\affiliation{Fakult\"{a}t Physik, Technische Universit\"{a}t Dortmund, D-44221
Dortmund, Germany.}

\author{Ashok Ajoy}

\affiliation{Fakult\"{a}t Physik, Technische Universit\"{a}t Dortmund, D-44221
Dortmund, Germany.}

\affiliation{Birla Institute of Technology and Science - Pilani, Zuarinagar, Goa
- 403726, India.}

\affiliation{NMR Research Centre, Indian Institute of Science, Bangalore - 560012,
India.}

\author{Xinhua Peng}

\affiliation{Fakult\"{a}t Physik, Technische Universit\"{a}t Dortmund, D-44221
Dortmund, Germany.}

\affiliation{Hefei National Laboratory for Physical Sciences at Microscale and
Department of Modern Physics, University of Science and Technology
of China, Hefei, Anhui 230026, People\textquoteright{}s Republic of
China }

\author{Dieter Suter}

\email{Dieter.Suter@tu-dortmund.de}

\affiliation{Fakult\"{a}t Physik, Technische Universit\"{a}t Dortmund, D-44221
Dortmund, Germany.}


\begin{abstract}
Avoiding the loss of coherence of quantum mechanical states is an
important prerequisite for quantum information processing. Dynamical
decoupling (DD) is one of the most effective experimental methods
for maintaining coherence, especially when one can access only the
qubit-system and not its environment (bath). It involves the application
of pulses to the system whose net effect is a reversal of the system-environment
interaction. In any real system, however, the environment is not static,
and therefore the reversal of the system-environment interaction becomes
imperfect if the spacing between refocusing pulses becomes comparable
to or longer than the correlation time of the environment. The efficiency
of the refocusing improves therefore if the spacing between the pulses
is reduced. Here, we quantify the efficiency of different DD sequences
in preserving different quantum states. We use $^{13}$C nuclear spins
as qubits and an environment of $^{1}$H nuclear spins as the environment,
which couples to the qubit via magnetic dipole-dipole couplings. Strong
dipole-dipole couplings between the proton spins result in a rapidly
fluctuating environment with a correlation time of the order of 100
$\mu$s. Our experimental results show that short delays between the
pulses yield better performance if they are compared with the bath
correlation time. However, as the pulse spacing becomes shorter than
the bath correlation time, an optimum is reached. For even shorter
delays, the pulse imperfections dominate over the decoherence losses
and cause the quantum state to decay. 
\end{abstract}
\maketitle

\section{Introduction}

%
{}Quantum mechanical systems have an enormous potential for realizing
information processing devices that are qualitatively more powerful
than systems based on classical physics \cite{Nielsen00}. The main
requirement for quantum information processing (QIP) is that the system
evolves according to the Schrödinger equation, under the influence
of a Hamiltonian that is under precise experimental control. However,
no system is completely isolated, and disturbances from its surrounding
environment (bath) spoil the quantum identity of the system. This
process is often called decoherence \cite{Zurek2003} and limits the
time scale over which quantum information can be retained and the
distance over which it can be transmitted \cite{Chiara2005,Allcock2009,Alvarez2010}.
Reducing the effects of decoherence is therefore one of the main requirements
for reliable quantum information processing. Several protocols have
been developed for quantum error correction \cite{Preskill1998,Knill2005};
however, they prove advantageous only for low levels of environmental
noise.

In all existing experimental architectures for QIP the noise background
is too large, and this limits the applicability of these protocols.
A promising technique for reducing the noise to a level where error-correcting
codes can take over is called Dynamical Decoupling (DD) \cite{viola_dynamical_1999,Yang2010}.
It aims to reduce the interaction of the system with the environment
through control operations acting only on the system. It requires
relatively modest resources, since it requires no overhead of information
encoding, measurements or feedback.

Although the mathematical framework of dynamical decoupling was introduced
fairly recently \cite{viola_dynamical_1999}, the six-decade old Hahn
NMR spin-echo experiment \cite{Hahn1950} can be considered as the
earliest and simplest implementation of this method. It consists of
the application of a $\pi$-pulse to a spin qubit ensemble, at time
$\tau$ after the spins were left to undergo Larmor precession in
a magnetic field. This effectively reverses a pure dephasing system-environment
(SE) interaction, i.e. one that does not cause a net exchange of energy
between the system and the bath. The combined effect of the evolution
before the refocusing pulse and a second period of the same duration
after the pulse vanishes. %
{}Physically, the dephasing and rephasing of the spins can be observed
as an apparent decay of the average magnetization in the system and
a subsequent increase after the refocusing pulse (a spin echo).

The Hahn-echo can (for an ideal pulse) completely eliminate the interaction
with the environment, provided it is time-invariant. In practice,
this is often not the case, and a change in the environment reduces
the refocusing efficiency \cite{Hahn1950,Carr1954}. To reduce the
problems due to a time-dependent environment, Carr and Purcell suggested
to replace the single pulse of the Hahn echo by a sequence of pulses
at shorter intervals (the CP sequence) \cite{Carr1954}, thus reducing
the changes in the environment between successive pulses. For sufficiently
short pulse intervals, elimination of system-environment interactions
became possible even in a time-dependent environment. However, the
increased number of pulses led to another problem: if the refocusing
pulses are not perfect, they actually become a source of decoherence
(and thus signal loss) instead of eliminating it. This problem was
significantly reduced by a simple modification of the CP sequence:
if the rotation axis of the refocusing pulses is parallel to the initial
spin orientation, the effect of pulse errors is significantly reduced
over a cycle \cite{Meiboom1958}. This is known in literature as the
CPMG sequence.

In the context of QIP, there has been renewed effort in eliminating
the effects of the system-environment interaction that lead to the
loss of quantum information. For this, it is often important that
the refocusing reduces the effect of the system-environment interactions
by several orders of magnitude. In addition, the effect of pulse errors
must be minimized, and the sequence has to work for all possible initial
states of the system. Several pulse sequences that achieve this were
introduced \cite{Maudsley1986,Gullion1990,viola_dynamical_1999,Viola2003},
which consist of periodic sequences of pulses; they were thus called
periodic DD (PDD). By design, they allow one to decouple the system
from the environment for a general SE interaction, i.e. one that causes
dephasing as well as dissipation.

Experimentally, DD is achieved by iteratively applying to the system
a series of stroboscopic control pulses in cycles of period $\tau_{c}$.
Over that period, the time-averaged SE interaction Hamiltonian vanishes.
The time average over $\tau_{c}$ can be calculated using average
Hamiltonian theory \cite{Haeberlen1976}. If the average Hamiltonians
are calculated by a series expansion, such as the Magnus expansion,
improving the pulse sequence usually corresponds to progressively
eliminating higher order terms in the expansion. Khodjasteh and Lidar
\cite{khodjasteh_fault-tolerant_2005} introduced concatenated DD
(CDD) as a scheme that recursively generates higher order DD sequences
for this purpose. Here, the lowest level of concatenation is a PDD
sequence. The improvement achieved by concatenation comes at the expense
of an exponential growth \cite{khodjasteh_fault-tolerant_2005} in
the number of applied control pulses. In contrast, for the case of
a pure dephasing or pure dissipative interaction Hamiltonian \cite{Yang2008,Lee2008},
Uhrig developed a sequence (UDD) \cite{Uhrig2007} that reduces higher
orders in the Magnus expansion with only a linear overhead in the
number of pulses. Unlike other DD sequences, in the UDD sequence the
delay between successive pulses is not equal, i.e. the pulses are
not equidistant. In the limit of a two-pulse cycle, UDD reduces to
the CPMG sequence. Recent proposals of DD sequences that are a hybrid
between UDD and CDD are predicted to improve DD performance of previous
methods \cite{Uhrig2009,west_near-optimal_2010}.

The UDD sequence was tested %
{}on ion traps \cite{biercuk_optimized_2009,Biercuk2009}, electron
paramagnetic resonance \cite{Du2009} and liquid-state NMR \cite{Jenista2009},
and found to outperform equidistant pulse sequences, in particular
CPMG, for environments with a high-frequency or strong cutoff. CDD
sequences were recently tested in solid-state NMR \cite{west_high_2009}.
However, while some sequences for particular environmental noises
were tested, a comparison between sequences for different kinds of
environments is still missing. Most of the sequences were designed
assuming ideal pulses and some of them predict to compensate pulse
imperfection. However an experimental test of this aspect is still
needed. In parallel to this work recent DD implementations on a qubit
interacting with a slowly fluctuating spin-bath were tested \cite{Bluhm2010,Barthel2010,Naydenov2010,Lange2010,Ryan2010}.

Other questions relate to the optimal cycle time: It is theoretically
predicted and experimentally demonstrated that sequences that reduce
higher order terms of the Magnus expansion perform better than low
order sequences for slow motion environments with high-frequency or
strong cutoff, when the bath correlation time $\tau_{B}$ is longer
than the sequence time $\tau_{c}$ (cycle time). However the strength
and duration of control-pulses are limited by hardware, yielding a
minimum for the achievable DD cycle time. As some examples on this
direction, Viola and Knill proposed a general method for DD with bounded
controls \cite{Viola2003}. Khodjasteh and Lidar, keeping the delay
between pulses constant, predicted an optimal CDD order for reducing
decoherence \cite{khodjasteh_performance_2007,Ng2009}. Biercuk \emph{et
al.} \cite{biercuk_optimized_2009,Biercuk2009} needed to consider
the finite length of pulses in their simulations, assuming them perfect
but producing a spin-lock during their application times, in order
to fit them to the experiments. Additionally Hodgson \emph{et al}.
\cite{Hodgson2010}, while assuming instantaneous perfect pulses,
theoretically analyzed DD performance constraining also the delay
between pulses. They set a lower limit to the delays making them larger
than the pulse duration to satisfy the instantaneous pulse approximation
in their theoretical model for experimental conditions. When the delays
are strongly constrained, they predict that DD protocols like CDD
or UDD, which are designed to improve the performance of lower DD
orders if the regime of arbitrarily small pulse separations is achievable,
in general lose their advantages. However, experiments are missing
in order to demonstrate these predictions and very little is known
about the performance of DD sequences under conditions where the cycle
times are comparable to or longer than the bath correlation times.
Recently Pryadko and Quiroz approached this regime, but only for the
extreme case of a Markovian environment \cite{Pryadko2009}%
{}%
{}.

While the finite length of pulses limits the minimum cycle time reducing
the maximal achievable DD performance, their imperfections also contribute
to reducing it. It is well know that CPMG-like sequences are too sensitive
to the initial state when pulse errors are considered \cite{Maudsley1986,Gullion1990}.
A comparisons between the CPMG and UDD sensitivity against pulse errors
was performed in Ref. \cite{Biercuk2009}. Overall UDD was shown to
be more robust against flip angle errors and static offset errors,
with the exception that CPMG is more robust for initial states longitudinal
to the control pulses. But both of them are too asymmetric against
initial state directions. In general, while some DD sequences were
developed to compensate flip-angle errors and to have a performance
more symmetric against initial conditions, an extensive study of their
performance from a QIP perspective is still missing and additionally
is not done for CDD sequences. For example, an optimal cycle time
when considering imperfect finite pulses was predicted by Khodjasteh
and Lidar \cite{khodjasteh_performance_2007}.

In this article, we compare experimentally the performance of different
DD sequences on a spin-based solid-state system where the cycle time
$\tau_{c}$ is comparable to or longer than the correlation time $\tau_{B}$
of the environment. Here, the spin(qubit)-system interacts with a
spin-bath where the spectral density of the bath is given by a normal
(Gaussian) distribution. This kind of systems, typical in NMR \cite{Abragam},
are encountered in a wide range of solid-state systems, as for example
electron spins in diamonds \cite{Naydenov2010,Lange2010,Ryan2010},
electron spins in quantum dots \cite{Hanson2007,Bluhm2010,Barthel2010}
and donors in silicon \cite{Kane1998,Morton2008} which appear to
be promising candidates for future QIP implementations. In particular
we consider the case where the interaction with the bath is weak compared
with the intra-bath interaction. For one side the latter point complement
and distinguish our work from the recent submitted articles \cite{Bluhm2010,Barthel2010,Naydenov2010,Lange2010,Ryan2010}.
For the other the aim of our work is a comprehensively and detailed
comparison of the performance of different sequences considering different
initial states. We find how the performance of the DD sequences depends
on the initial state of the qubit ensemble with respect to the rotation
axis of control pulses with finite precision. When they are in the
same direction, the CPMG sequence is the best DD sequence for reducing
decoherence -- i.e. it maintains the state of the ensemble for the
longest time. However, if the initial state of the ensemble is not
known, we find that concatenated dynamical decoupling (CDD) provides
the best overall performance. Stated equivalently, the CDD scheme
provides the best overall minimization of the environmentally driven
quantum mechanical evolution of the system. Additionally we experimentally
demonstrate and quantify the predicted optimal delay times for maximizing
the performance of the respective DD sequences. This implies that
pulse errors are a limiting factor that must be reduced to improve
DD performances. In general our results complement some of the previous
findings and predictions for some of the experimentally tested DD
sequences and provide new results for untested ones. One of the main
message is that a fair comparison of the performance of DD sequences
should use a constant average number of pulses per unit time.

This paper is organized as follows. Section \ref{sec:system} describes
the qubit and bath system used in our experiment, and the mechanisms
of coupling between them. In section \ref{sec:Dynamical-decoupling}
we give a brief summary of dynamical decoupling and a description
of the tested sequences -- the Hahn Echo, CPMG, PDD, CDD and UDD.
Our limited choice of sequences includes those most accepted by the
QIP community and allows us to discuss the most important points.
Section \ref{sec:NMR-dynamical-decoupling} contains the experimental
results and their analysis. In section \ref{sec:ComparisonsOptimal-choices}
we compare the various DD sequences under the same conditions. In
the last section we draw some conclusions.%
{}%
{}

\section{The system\label{sec:The-system}}

\label{sec:system} Our system consists of a spin 1/2 (qubit) in a
strong magnetic field oriented along the $z$-axis, interacting with
a bath consisting of a different type of spins 1/2. The total Hamiltonian
in the laboratory frame is \begin{equation}
\widehat{\mathcal{H}}^{L}=\widehat{\mathcal{H}}_{S}^{L}+\widehat{\mathcal{H}}_{SE}^{L}+\widehat{\mathcal{H}}_{E}^{L},\label{eq:H}\end{equation}
 where $\widehat{\mathcal{H}}_{S}^{L}$ is the system Hamiltonian,
$\widehat{\mathcal{H}}_{E}^{L}$ is the environment Hamiltonian and
$\widehat{\mathcal{H}}_{SE}^{L}$ is the system-environment interaction
Hamiltonian: \begin{align}
\widehat{\mathcal{H}}_{S}^{L} & =\omega_{S}\hat{S}_{z},\label{eq:HSL}\\
\widehat{\mathcal{H}}_{SE}^{L} & =\hat{S}_{z}\sum_{j}b_{Sj}\hat{I}_{z}^{j},\label{eq:HSEL}\\
\widehat{\mathcal{H}}_{E}^{L} & =\omega_{I}\sum_{j}\hat{I}_{z}^{j}+\sum_{i<j}d_{ij}\left[2\hat{I}_{z}^{i}\hat{I}_{z}^{j}-(\hat{I}_{x}^{i}\hat{I}_{x}^{j}+\hat{I}_{y}^{i}\hat{I}_{y}^{j})\right],\label{eq:HEL}\end{align}
 where $\hat{S}$ is the spin operator of the system qubit, the spin
operators $\hat{I}_{x}^{j},\hat{I}_{y}^{j}\mbox{ and }\hat{I}_{z}^{j}$
act on the $j^{th}$ bath spin, $\omega_{S}$ and $\omega_{I}$ are
the Zeeman frequencies of the system spin and the bath spins respectively,
$b_{Sj}$ and $d_{ij}$ are the coupling constants, and we use frequency
units ($\hbar=1$). In solids, the spin-spin interaction is dominated
by the dipolar interaction \cite{Abragam}. Since $S$ and $I$ are
different types of nuclei, it is possible to neglect the terms of
the dipolar coupling Hamiltonian that do not commute with the strong
Zeeman interaction because $\left|b_{Sj}\right|/\left|\omega_{S}-\omega_{I}\right|\ll10^{-4}$
\cite{Abragam}. The remaining terms have the Ising form (\ref{eq:HSEL}).
Similarly, the homonuclear interaction between the bath spins is truncated
to those terms that commute with the total Zeeman coupling, which
we assume to be identical for all bath spins.

In the high-temperature thermal equilibrium \cite{Abragam}, the density
operator of the system spin is \begin{equation}
\hat{\rho}_{S,\textrm{eq.}}\propto\hat{S}_{z},\end{equation}
 where we consider only the system ($S$-spin) part of the total Hilbert
space. We also neglect the part proportional to the unit operator,
which does not evolve in time and does not contribute to the observable
signal.

To generate the initial state for our DD measurements, we rotate the
thermal state to the $xy$-plane by applying a $\pi/2$ pulse. The
resulting state is \begin{equation}
\hat{\rho}_{S}(0)\propto\hat{S}_{\left\{ {x\atop y}\right\} }.\end{equation}
 For an isolated spin system this magnetization precesses indefinitely
around the static magnetic field at the Zeeman frequency $\omega_{S}$.

Taking the system-environment interaction into account, the effect
of the coupling operator $\widehat{\mathcal{H}}_{SE}^{L}$ is the
generation of product terms of the form $\hat{S}_{\pm}\hat{I}_{z}^{j}$
in the density operator, correlating the system with the environment.
Since we only observe the system part of the total Hilbert space,
we effectively project the correlated system onto this subspace,\begin{equation}
\hat{\rho}_{S}=\mathrm{Tr}_{I}\left\{ \hat{\rho}_{tot}\right\} ,\end{equation}
 where $\mathrm{Tr}_{I}$ represents the partial trace over the environmental
degrees of freedom and $\hat{\rho}_{tot}$ represents the density
operator of system plus environment. The result of this projection
corresponds to a loss of coherence by dephasing. This free evolution
of the system under the SE interaction is called the free induction
decay (FID) in NMR terminology. The decay process of the system state
is usually called relaxation in NMR terminology, or decoherence in
quantum information.

In the following, we will describe the dynamics of the system in a
rotating frame of reference \cite{Abragam}: The system rotates at
the (angular) frequency $\omega_{S}$ around the $z$-axis and the
environment at $\omega_{I}$. As a result, the rotating frame Hamiltonian
becomes\begin{equation}
\widehat{\mathcal{H}}_{f}=\widehat{\mathcal{H}}_{S}+\widehat{\mathcal{H}}_{SE}+\widehat{\mathcal{H}}_{E},\label{eq:Hfreeevolution}\end{equation}
 where \begin{align}
\widehat{\mathcal{H}}_{S} & =\widehat{\mathcal{H}}_{S}^{L}-\omega_{S}\hat{S}_{z}=\hat{0},\label{eq:HS}\\
\widehat{\mathcal{H}}_{SE} & =\hat{S}_{z}\sum_{j}b_{Sj}\hat{I}_{z}^{j},\label{eq:HSE}\end{align}
 \begin{multline}
\widehat{\mathcal{H}}_{E}=\widehat{\mathcal{H}}_{E}^{L}-\omega_{I}\sum_{j}\hat{I}_{z}^{j}\\
=\sum_{i<j}d_{ij}\left[2\hat{I}_{z}^{i}\hat{I}_{z}^{j}-(\hat{I}_{x}^{i}\hat{I}_{x}^{j}+\hat{I}_{y}^{i}\hat{I}_{y}^{j})\right].\label{eq:HE}\end{multline}
 This transformation is exact, since the Zeeman terms commute with
all other terms in the Hamiltonian as well as with the equilibrium
density operator.

The effect of the environment-Hamiltonian $\mathcal{\widehat{H}}_{E}$
on the evolution of the system may be discussed in an interaction
representation with respect to the evolution of the isolated environment:
the system-environment Hamiltonian then becomes\begin{align}
\mathcal{\widehat{H}}_{SE}^{(E)}\left(t\right) & =e^{-i\mathcal{\widehat{H}}_{E}t}\mathcal{\widehat{H}}_{SE}e^{i\mathcal{\widehat{H}}_{E}t}\nonumber \\
 & =\hat{S}_{z}e^{-i\mathcal{\widehat{H}}_{E}t}\left(\sum_{j}b_{Sj}\hat{I}_{z}^{j}\right)e^{i\mathcal{\widehat{H}}_{E}t}.\label{eq:HSE-time-dependent}\end{align}
 %
{}%
{} Since $\mathcal{\widehat{H}}_{E}$ does not commute with $\mathcal{\widehat{H}}_{SE}$,
the effective system-environment interaction $\mathcal{\widehat{H}}_{SE}^{\left(E\right)}$
becomes time-dependent: the system experiences a coupling to the environment
that fluctuates. %
{}The correlation time $\tau_{B}$ of the time-dependent spin-bath operators
$\hat{I}_{z}^{j}(t)=e^{-i\mathcal{\widehat{H}}_{E}t}\hat{I}_{z}^{j}e^{i\mathcal{\widehat{H}}_{E}t}$%
{} is defined by the decay to $1/e$ of the correlation function \begin{equation}
i_{z}^{j}(t)=\frac{\mathrm{Tr}\left\{ \hat{I}_{z}^{j}(0)\hat{I}_{z}^{j}(t)\right\} }{\mathrm{Tr}\left\{ \hat{I}_{z}^{j}(0)\hat{I}_{z}^{j}(0)\right\} }.\label{eq:izcorrfunc}\end{equation}
 Considering that all the bath spins are equivalent, the latter correlation
funtions are identical for every $j$ and share the same correlation
time $\tau_{B}$%
{}.

\section{Dynamical decoupling\label{sec:Dynamical-decoupling}}

\subsection{Notation}

The aim of dynamical decoupling is the reduction of the interaction
of the qubit system with the environment, thus retaining the quantum
information for as long as possible. In the context of DD, it is assumed
that it is possible to apply arbitrary single-qubit operations to
the system qubit, but that it is not possible to control the environment.
One thus applies to the system short, strong pulses, whose effect
can be described as a refocusing of the system-environment interaction
by the control Hamiltonians $\widehat{\mathcal{H}}_{{\rm \textrm{C}(S)}}(t)$
\cite{viola_dynamical_1999,Yang2010}.

\begin{figure}
\includegraphics[bb=17bp 479bp 530bp 705bp,scale=0.4]{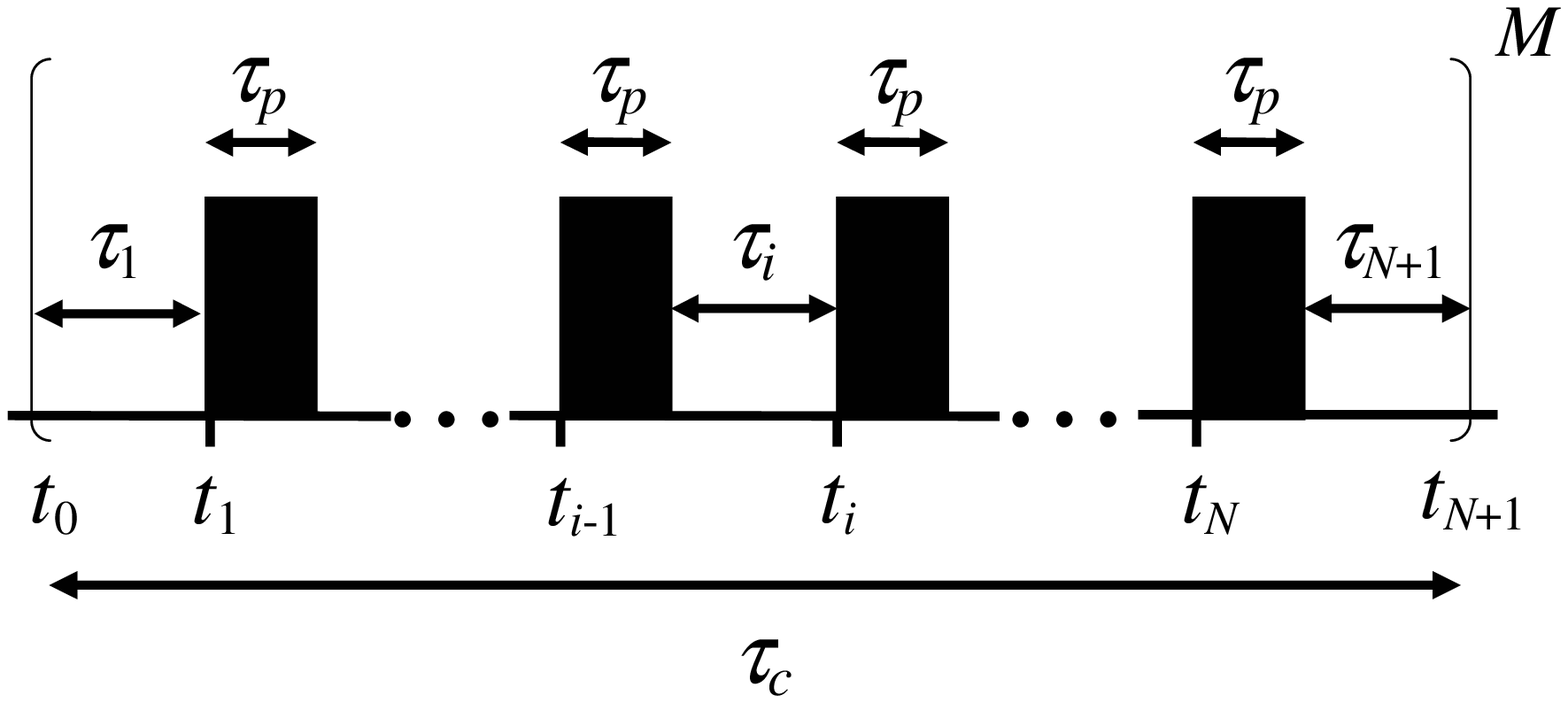}

\caption{Schematic representation of dynamical decoupling. The solid boxes
represents the control pulses .}

\label{Flo:DDScheme} 
\end{figure}

Let us refer to Fig. \ref{Flo:DDScheme} and consider a single cycle
of the sequence having a period $\tau_{c}$. In the rotating frame,
the operator that describes the evolution of the total system from
0 to $\tau_{c}$ is \begin{equation}
\hat{U}\left(\tau_{c}\right)=\hat{U}_{f}\left(\tau_{N+1}\right)\prod_{i=1}^{N}\hat{U}_{C}^{i}\left(\tau_{p}\right)\hat{U}_{f}\left(\tau_{i}\right),\label{eq:TotalU}\end{equation}
 where from Eq. (\ref{eq:Hfreeevolution}) the free evolution operator
is \begin{eqnarray}
\hat{U}_{f}\left(t\right) & = & \exp\left\{ -\mbox{i}\widehat{\mathcal{H}}_{f}t\right\} \end{eqnarray}
 and the control evolution operators that act during the time $\tau_{p}$
is\begin{eqnarray}
\hat{U}_{C}^{i}\left(\tau_{p}\right) & = & T\exp\left\{ -i\int_{0}^{\tau_{p}}dt^{\prime}\left(\widehat{\mathcal{H}}_{f}+\widehat{\mathcal{H}}_{{\rm \textrm{C}(S)}}^{i}(t^{\prime})\right)\right\} \label{eq:UC}\end{eqnarray}
 with $T$ the Dyson time-ordering operator \cite{dyson0,dyson1}.
We assume that the free evolution Hamiltonian is constant, while the
control Hamiltonian $\widehat{\mathcal{H}}_{{\rm \textrm{C}(S)}}^{i}(t)$
is constant during $\tau_{p}$ but changes for different $i$. The
delay times between the control Hamiltonians are $\tau_{i}=t_{i}-\left(t_{i-1}+\tau_{p}\right)$
for $i=2,..,N+1$ and $\tau_{1}=t_{1}-t_{0}$, where $t_{0}=0$, $t_{N+1}=\tau_{c}$,
and $t_{i}$ represents the time at which the $i^{th}$ control operation
starts. Figure \ref{Flo:DDScheme} shows a graphical representation
of these definitions. %

Like any unitary evolution, the total propagator can be written as
the exponential of a Hermitian operator, \begin{equation}
\hat{U}\left(t\right)=e^{-i\widehat{\mathcal{H}}_{eff}t}.\end{equation}
 Using average Hamiltonian theory \cite{Haeberlen1976} we can calculate
the effective Hamiltonian $\widehat{\mathcal{H}}_{eff}$ as a series
expansion, \begin{align}
\widehat{\mathcal{H}}_{eff} & =\widehat{\mathcal{H}}^{(0)}+\widehat{\mathcal{H}}^{(1)}+\widehat{\mathcal{H}}^{(2)}+...=\sum_{n=0}^{\infty}\widehat{\mathcal{H}}^{(n)}.\end{align}
 The zero order term $\widehat{\mathcal{H}}^{(0)}$ is given by the
time integral of the total Hamiltonian from time 0 to $\tau_{c}$.
An ideal DD sequence makes $\widehat{\mathcal{H}}^{(0)}=\widehat{\mathcal{H}}_{E}$
, i.e. for ideal pulses, the interaction Hamiltonian vanishes to zeroth
order.%
{} %
{}%
{}In the Magnus expansion \cite{Magnus1954}, higher order terms are
proportional to increasing powers of $\tau_{c}/\tau_{B}$, since we
assume that the environment is weakly coupled to the system ($b_{Sj}\tau_{B}\ll1$)
and in consequence $\tau_{B}$ is the dominant time-scale \cite{khodjasteh_performance_2007}.%
{}

If the basic cycle is iterated $M$ times (see Fig. \ref{Flo:DDScheme}),
the total evolution operator becomes \begin{equation}
\hat{U}\left(t=M\tau_{c}\right)=\left[\hat{U}\left(\tau_{c}\right)\right]^{M}.\end{equation}

\subsection{Ideal and real pulses}

The usual approximation of hard pulses -- having a radio-frequency
field $\omega_{p}\gg b_{Sj}$ and duration $\tau_{p}\ll d_{i,j}^{-1},b_{Sj}^{-1}$
implies that we can neglect the free precession Hamiltonian and Eq.
(\ref{eq:UC}) simplifies to \begin{equation}
\hat{U}_{C}^{i}\left(\tau_{p}\right)=\exp\left\{ -\mbox{i}\hat{S}_{u}\theta_{p}\right\} \end{equation}
 in the rotating frame, where $u=x,y,z$ and $\theta_{p}=\omega_{p}\tau_{p}$
is the rotation angle around the $u$ axis. In what follows, we shall
denote perfect instantaneous $\pi$-pulses along $x$ and $y$ by
$\hat{X}=\exp\left\{ -i\hat{S}_{x}\pi\right\} $ and $\hat{Y}=\exp\left\{ -i\hat{S}_{y}\pi\right\} $
respectively, and a free evolution of duration $\tau$ by $f_{\tau}$.
%
{}

To take the effect of non-ideal pulses into account, one needs to
consider errors in the axis and angle of rotation. We write the resulting
control propagator as the product of the ideal pulse rotation times
an error rotation $\exp\left\{ -\mathrm{i}\hat{S}_{e_{i}}\theta_{i,e}\right\} $:
\begin{equation}
\hat{U}_{C}^{i}\left(\tau_{p}\right)=\exp\left\{ -\mathrm{i}\hat{S}_{e_{i}}\theta_{i,e}\right\} \exp\left\{ -\mbox{i}\hat{S}_{u_{i}}\theta_{p}\right\} .\end{equation}
 The total evolution operator is thus\begin{equation}
\hat{U}\left(\tau_{C}\right)=\hat{U}_{f_{N+1}}^{\prime}\left(\tau_{N+1},\tau_{p}\right)\prod_{i=1}^{N}\hat{U}_{C}^{i}\left(0\right)\hat{U}_{f_{i}}^{\prime}\left(\tau_{i},\tau_{p}\right),\label{eq:modEvol}\end{equation}
 where the evolution operators\begin{equation}
\hat{U}_{f_{i}}^{\prime}\left(\tau_{i},\tau_{p}\right)=\hat{U}_{f}\left(\tau_{i}\right)\exp\left\{ -\mathrm{i}\hat{S}_{e_{i}}\theta_{i,e}\right\} \label{eq:EffectiveUf}\end{equation}
 represent a modified free evolution. Note that $\hat{U}_{f_{1}}^{\prime}\left(\tau_{i},\tau_{p}\right)=\hat{U}_{f}\left(\tau_{1}\right).$

The zero order average Hamiltonian of the free evolution periods (\ref{eq:EffectiveUf})
for non-perfect pulses is equivalent to interactions of the general
form\begin{multline}
\widehat{\mathcal{H}}_{SE}^{npp}=a_{x}\hat{S}_{x}+a_{y}\hat{S}_{y}+a_{z}\hat{S}_{z}\\
+\sum_{j}\left(b_{Sj}^{x}\hat{S}_{x}+b_{Sj}^{y}\hat{S}_{y}+b_{Sj}^{z}\hat{S}_{z}\right)\hat{I}_{z}^{j}\\
=\sum_{u=x,y,z}\hat{S}_{u}\sum_{j}\left(a_{u}+b_{Sj}^{u}\hat{I}_{z}^{j}\right),\label{eq:GSEI}\end{multline}
 where $a_{u}$ and $b_{Sj}^{u}$ give the renormalized offsets and
couplings respectively, which include the errors of the control Hamiltonians.
This picture can also consider errors of control pulses when $\tau_{p}$
is comparable with the inverse couplings of the free evolution Hamiltonian.

We now discuss some DD schemes that refocus the system-environment
interaction. In all these cases, we assume that the system is initially
prepared in a coherent superposition of the computational basis states.
We will refer to the initial state of the qubit as $\hat{S}_{x}$,
$\hat{S}_{y}$ or $\hat{S}_{z}$, as shown in Fig. \ref{Flo:NMR_seq}(a).

\subsection{Hahn echo}

\begin{figure}
\includegraphics[bb=38bp 56bp 580bp 698bp,clip,scale=0.35]{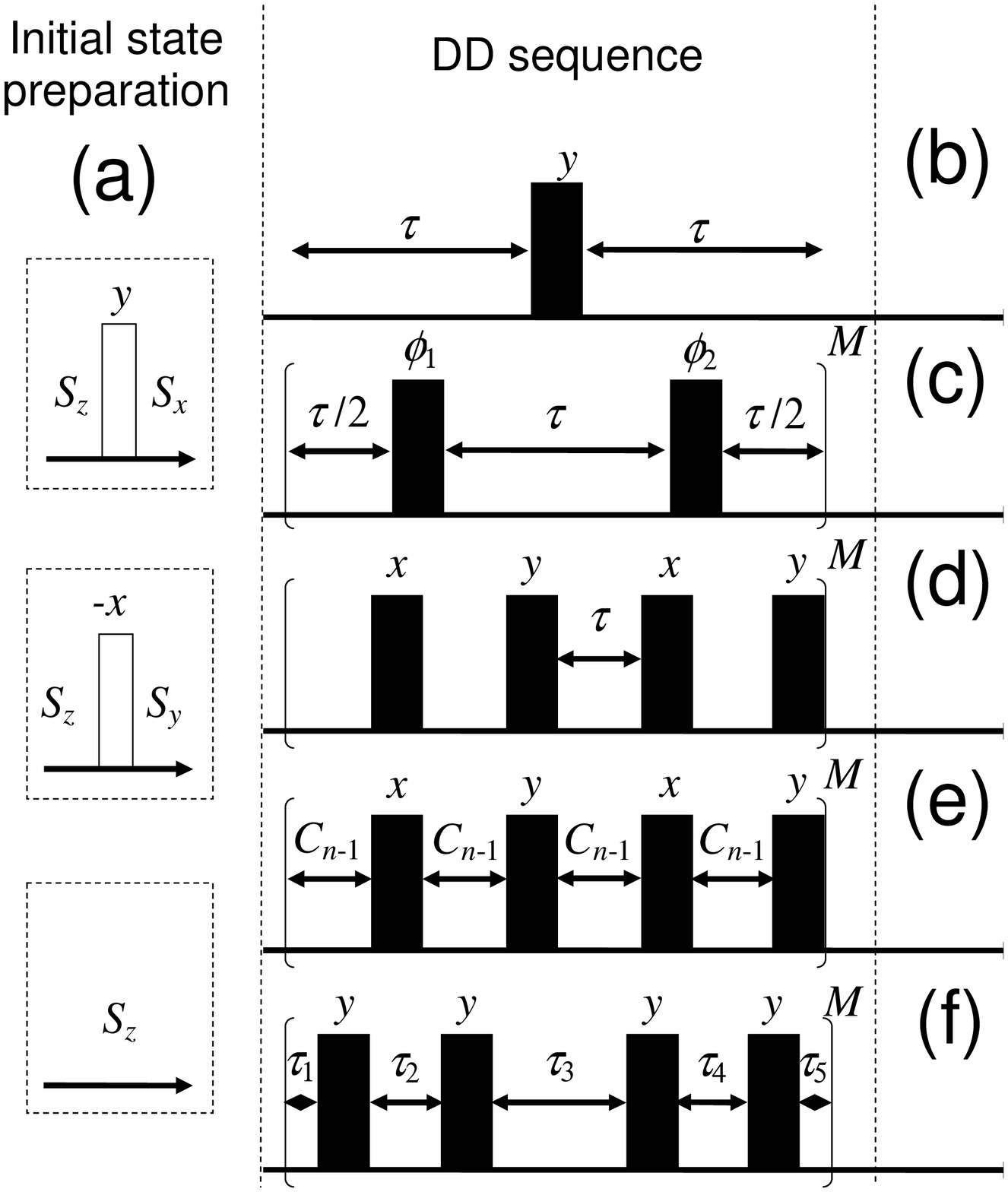}

\caption{Schemes of dynamical decoupling pulse sequences. Empty and solid rectangles
represent $\pi/2$ and $\pi$ pulses respectively. $M$ represents
the number of iterations of the cycle. (a) Initial state preparation
before application of the DD sequence. (b) Hahn spin-echo sequence.
(c) CPMG ($\phi_{2}=\phi_{1}$) and CPMG-2 ($\phi_{2}=\phi_{1}+\pi$)
sequences. (d) PDD sequence. (e) CDD sequence of order $n$, CDD$_{n}=C_{n}$.
(f) UDD sequence scheme with 4 pulses, i.e. UDD of order 4, UDD$_{4}$. }

\label{Flo:NMR_seq} 
\end{figure}

The Hahn spin-echo experiment \cite{Hahn1950} is the pioneer dynamical
decoupling method and the building block for newer DD proposals. It
consists of the application of a $\pi$-pulse to the $S$ spin along
an axis (say $y$) transverse to the static field $B_{0}$ at time
$\tau$ causing an echo at time $2\tau$ {[}Fig. \ref{Flo:NMR_seq}(b){]}.
The total evolution operator can be summarized as $f_{\tau}\hat{Y}f_{\tau}$
where the total time (assuming a delta-function pulse) is $2\tau$.
As a consequence, the zero-order average Hamiltonian is, \begin{equation}
\widehat{\mathcal{H}}_{\mathrm{Hahn}}^{(0)}=\frac{1}{2\tau}\int_{0}^{2\tau}dt^{\prime}\widehat{\mathcal{H}}\left(t^{\prime}\right)=\frac{\left(\tau\widehat{\mathcal{H}}_{SE}-\tau\widehat{\mathcal{H}}_{SE}\right)}{2\tau}=0.\end{equation}
 The resulting system evolution operator approaches the identity to
within $\mathcal{O}\left(\left(\tau_{c}/\tau_{B}\right)^{2}\right)$%
{}. Thus if $\tau_{c}\ll\tau_{B}$, a perfect echo (time reversion)
is achieved at the total evolution time $t=2\tau=\tau_{c}$. When
$\tau_{c}$ is comparable to or longer than $\tau_{B}$, the echo
decays due to the higher order terms.%
{}

\subsection{Carr-Purcell (CP) and Carr-Purcell-Meiboom-Gill (CPMG)}

To avoid the decay of the echo due to the finite correlation time
of the environment, Carr and Purcell \cite{Carr1954} reduced the
cycle time by splitting the total time into shorter segments of equal
length, and a refocusing pulse in the middle of each segment.

Figure \ref{Flo:NMR_seq}(c) shows the pulse sequence for an initial
condition of $\hat{S}_{y}$ with $\phi_{1}=\phi_{2}=y$. The resulting
evolution operator is $f_{\tau/2}\hat{Y}f_{\tau}\hat{Y}f_{\tau/2}$.
Later on, Meiboom and Gill \cite{Meiboom1958} suggested to shift
the phase of the refocusing pulses by $\pi/2$, so that the rotation
axis is the same as the orientation of the initial state. For perfect
pulses, both cases are equivalent, but only the CPMG version compensates
flip-angle errors of the refocusing pulses. For a flip-angle error
$\exp\left\{ -\mathrm{i}\hat{S}_{e_{i}}\theta_{i,e}\right\} =\exp\left\{ -\mathrm{i}\hat{S}_{y}\Delta\omega_{1}\tau_{p}\right\} $
for every $i$ in Eq. (\ref{eq:EffectiveUf}), the zero-order average
Hamiltonian is proportional to $\Delta\omega_{1}\hat{S}_{y}$. It
thus commutes with an initial condition along the $y$ axis, (the
CPMG case) and has no effect, but it causes an unwanted rotation of
an initial state $\propto\hat{S}_{x}$ (the CP case). In the following
we call this sequence with identical $\pi$-pulses CPMG.

An alternative sequence that also compensates flip-angle errors of
the refocusing pulses is shown in Fig. \ref{Flo:NMR_seq}(c), with
$\phi_{2}=\phi_{1}+\pi=-y$. For hard pulses and vanishing delays
between the pulses, the zero-order average Hamiltonian of this sequence
vanishes, for arbitrary flip-angle errors, and the first non-vanishing
term is of order $\tau_{c}/\tau_{B}$ and proportional to $\hat{S}_{x}$.
As a consequence, an initial condition proportional to $\hat{S}_{x}$
is less affected under this sequence. In what follows, we will call
this DD sequence CPMG-2.

The effect of pulse errors during CPMG and CPMG-2 on the spin dynamics
was studied in Refs. \cite{Franzoni2005,li_generating_2007,Franzoni2008,li_intrinsic_2008,dong_controlling_2008,Franzoni2010,Franzoni2010a},
and we will show some effects in the following sections.

\subsection{Periodic Dynamical decoupling (PDD)}

A sequence called XY-4 in the NMR community was proposed initially
to compensate the sensitivity of the CPMG-like sequences to non-perfect
pulses \cite{Maudsley1986,Gullion1990}. Later, it was found equivalent
to the shortest universal DD sequence that cancels the zero order
average Hamiltonian for a general SE interaction of the form (\ref{eq:GSEI})
\cite{viola_dynamical_1998,khodjasteh_performance_2007}. This sequence,
depicted in Fig. \ref{Flo:NMR_seq}(d) and called periodic dynamical
decoupling (PDD), has an evolution operator of the form $\hat{Y}f_{\tau}\hat{X}f_{\tau}\hat{Y}f_{\tau}\hat{X}f_{\tau}$.
Because it suppresses SE interactions of the form (\ref{eq:GSEI}),
it compensates errors of non-ideal pulses at the end of the cycle.
%
{}

\subsection{Concatenated Dynamical Decoupling (CDD)}

The concatenated DD (CDD) scheme \cite{khodjasteh_fault-tolerant_2005,khodjasteh_performance_2007}
recursively concatenates lower order sequences to effectively increase
the decoupling order. The CDD evolution operator for a recursion order
of $n$ is given by \begin{equation}
\mathrm{CDD}_{n}=C_{n}=\hat{Y}C_{n-1}\hat{X}C_{n-1}\hat{Y}C_{n-1}\hat{X}C_{n-1},\end{equation}
 where $C_{0}=f_{\tau}$ and $\mathrm{CDD}_{1}=\mathrm{PDD}$. Fig.
\ref{Flo:NMR_seq}(e) shows a general scheme for this process. Each
level of concatenation reduces the norm of the first non-vanishing
order term of the Magnus expansion of the previous level, provided
that the norm was small enough to begin with. The latter reduction
is at the expense of an extension of the cycle time by a factor of
four.

\subsection{Uhrig dynamical decoupling (UDD)}

Uhrig proposed a different approach to the goal of keeping a qubit
alive \cite{Uhrig2007,Uhrig2008}: For a given number $N$ of pulses
during a total time $\tau_{c}$, at what times should these pulses
be applied to minimize the effect of the system-environment interaction?
The solution he found for the times $t_{i}$ is\begin{equation}
t_{i}=\tau_{c}\sin^{2}\left[\frac{\pi i}{2\left(N+1\right)}\right],\end{equation}
 where $t_{N+1}=\tau_{c}$ is the cycle time and $t_{0}=0$ the starting
time. Defining $\tau_{i}=t_{i}-t_{i-1}$ the UDD evolution operator
for a sequence of $N$ pulses is \begin{equation}
\mathrm{UDD}_{N}=f_{\tau_{N+1}}\hat{Y}f_{\tau_{N}}\hat{Y}...\hat{Y}f_{\tau_{2}}\hat{Y}f_{\tau_{1}}\end{equation}
 and its schematic representation is given in Fig. \ref{Flo:NMR_seq}(f).
The CPMG sequence is the simplest UDD sequence of order $N=2$.

Cywinski \emph{et al.} explained the performance of the DD sequence
by finding its spectral filter for the bath-modes \cite{Cywinski2008}.
They found that the effect of the UDD pulse sequence leads to an efficient
spectral filter for slow motion bath-modes. It was shown that UDD
is the best sequence for reducing the SE interaction in the limit
of low-frequency noise \cite{Uhrig2007,Cywinski2008,Uhrig2008}. Rigorous
performance bounds for the UDD sequence were found by Uhrig and Lidar
in Ref. \cite{UhrigLidar2010}. %
{}

\section{Experimental results\label{sec:NMR-dynamical-decoupling}}

\subsection{System and environment }

Experiments were performed on a polycrystalline adamantane sample
using a home-built solid state NMR spectrometer with a $^{\text{1}}$H
resonance frequency of 300 MHz. The adamantane molecule contains two
nonequivalent carbon atoms. Under our conditions, they have similar
dynamics. Working with natural abundance (1.1 \%), the interaction
between the$^{13}$C-nuclear spins can be neglected. The main mechanism
for decoherence is the interaction with the proton spins. As discussed
in section \ref{sec:The-system}, this interaction (\ref{eq:HSE-time-dependent})
is not static, since the dipole-dipole couplings within the proton
bath cause flip-flops of the protons coupled to the carbon.

\begin{figure}
\includegraphics[bb=0bp 0bp 300bp 253bp,scale=0.8]{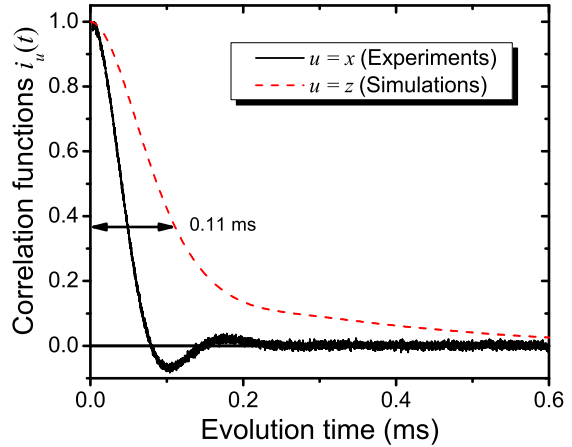}

\caption{(color online) Evolution of the normalized spin correlation functions
for the bath spins (protons). The solid line represents the proton
FID signal {[}$i_{x}\left(t\right)${]} and the dashed line the numerically
simulated $i_{z}\left(t\right)$.}

\label{Flo:ProtonFID} 
\end{figure}

Considering that all the proton spins $I$ are equivalent, we can
estimate the correlation time of the time-dependent SE interaction
(\ref{eq:HSE-time-dependent}) with the decay time of the correlation
function $i_{z}^{j}(t)$. While the correlation functions $i_{z}^{j}(t)$
of Eq. (\ref{eq:izcorrfunc}) cannot be measured directly because
we cannot address individual spins of the bath, we get a very good
estimate by measuring\begin{align}
i_{x}(t) & =\frac{\mathrm{Tr}\left\{ \hat{I}_{x}(0)\hat{I}_{x}(t)\right\} }{\mathrm{Tr}\left\{ \hat{I}_{x}(0)\hat{I}_{x}(0)\right\} },\label{eq:Ixcorrelation}\end{align}
 i.e. the proton free-induction decay (FID) (solid line in Fig. \ref{Flo:ProtonFID}).
The time evolutions in equation (\ref{eq:Ixcorrelation}) are determined
by the bath Hamiltonian $\mathcal{\widehat{H}}_{E}$, $\hat{I}_{x}(t)=e^{-i\mathcal{\widehat{H}}_{E}t}\hat{I_{x}}e^{i\mathcal{\widehat{H}}_{E}t}$
and $\hat{I}_{x}=\sum_{j}\hat{I}_{x}^{j}$. Simulating it with the
dipole-dipole Hamiltonian of Eq. (\ref{eq:HE}) and using the same
Hamiltonian for calculating $i_{z}^{j}(t)$,%
{} we find the correlation function represented with dashed line in
Fig. \ref{Flo:ProtonFID}. The spectral density of the bath is well
approximated by a normal (Gaussian) distribution and the system-environment
interaction is weak compared with the intra-bath interaction $\left(\left|b_{Sj}\right|\tau_{B}\lesssim1/3\right)$.

The $\text{\ensuremath{\pi}}$ pulses for DD were applied on resonance
with the $^{13}$C spins. Their radio-frequency (RF) field of $2\pi\times48$kHz
gives a $\text{\ensuremath{\pi}}$-pulse length of $\text{\ensuremath{\tau_{p}}=}10.4\mu\mbox{s}$.
The measured RF field inhomogeneity is about 10\%. We performed experiments
where the delay $\tau$ between successive DD pulses was varied from
$10\mu\mbox{s}$ to $200\mu\mbox{s}.$ We prepared the initial state
by using the sequences of Fig. \ref{Flo:NMR_seq}(a) and we measured
the survival probability%
{} of the magnetization \begin{equation}
s_{u}\left(t\right)=\frac{\text{Tr}\left\{ \hat{S}_{u}\left(0\right)\hat{S}_{u}\left(t\right)\right\} }{\text{Tr}\left\{ \hat{S}_{u}\left(0\right)\hat{S}_{u}\left(0\right)\right\} },\label{eq:SurvivalP}\end{equation}
 where $u=x,y,z$. The solid line of Fig. \ref{Flo:CarbonFIDHAHN}
shows the experimental observation of this survival probability from
an initial condition $\hat{S}_{x}$ under a free evolution ($^{13}$C
FID).

\subsection{Hahn echo}

\begin{figure}
\includegraphics[bb=0bp 0bp 301bp 253bp,scale=0.8]{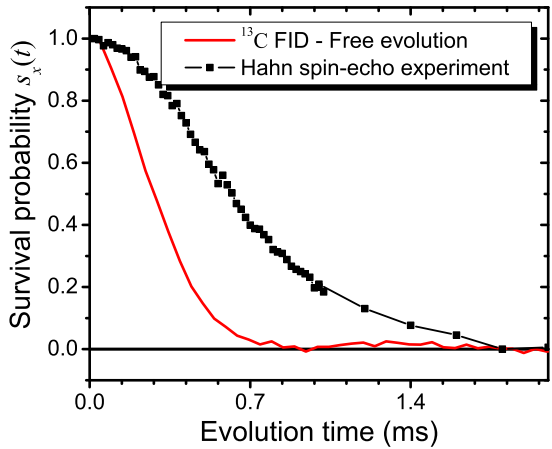}

\caption{(color online) Survival probability of the $S$-spin under free evolution
($^{13}$C FID) and after a Hahn echo sequence. An initial condition
$\hat{S}_{x}$ was prepared.}

\label{Flo:CarbonFIDHAHN} 
\end{figure}

As shown in Fig. \ref{Flo:CarbonFIDHAHN} , the decay of the $S$-spin
magnetization is reduced by the Hahn echo sequence. The results of
the Hahn echo are marked by square points. Compared to the free induction
decay, the decay rate is reduced approximately by a factor of $2$.

\subsection{CPMG}

\begin{figure}
\includegraphics[bb=10bp 10bp 250bp 315bp,clip]{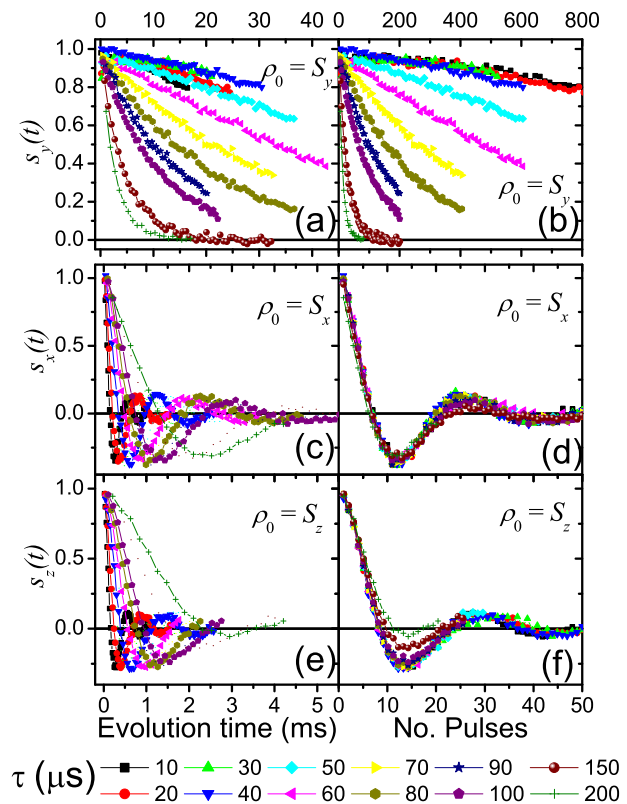}

\caption{(color online) Magnetization evolution for the CPMG sequence. From
top to bottom the initial conditions are $\hat{S}_{y}$, $\hat{S}_{x}$
and $\hat{S}_{z}$. The left-hand panels represent the $^{13}$C magnetization
as a function of the total evolution time while right-hand panels
show its evolution as a function of the number of applied pulses.
The legend at the bottom gives the delays $\tau$ between successive
pulses.}

\label{Flo:CPMG} 
\end{figure}

Figure \ref{Flo:CPMG} shows the experimental results of the CPMG
sequence of Fig. \ref{Flo:NMR_seq}(c) with $\phi_{1}=\phi_{2}=y$.
Different rows correspond to different initial conditions $\hat{S}_{x}$,
$\hat{S}_{y}$ and $\hat{S}_{z}$ of the $^{13}$C qubit. The left
hand panels show the survival probability (\ref{eq:SurvivalP}) as
a function of the total evolution time $t=M\tau_{c}$ (including the
pulses), and the right hand panels show the same data as a function
of the number of applied pulses.

The plots show that the decay of the survival probability depends
crucially on the initial state of the qubit; we shall henceforth refer
to the initial state in the direction of the DD pulses as the {}``longitudinal''
state, and the ones perpendicular to the pulses as the {}``transverse''
states. Flip-angle errors, which arise from inhomogeneous radio-frequency
fields, are usually the dominant imperfection in this type of experiments.
When the CPMG sequence is applied to a longitudinal initial condition,
flip-angle errors do not affect the performance of the decoupling,
since they are compensated over each cycle consisting of two pulses
\cite{Meiboom1958}. As a result, the decay rates for longitudinal
states are about an order of magnitude lower than for transverse initial
conditions. We also observe an unexpected oscillation pattern for
transverse initial states. These kind of strong asymmetries have been
reported in different samples, and have been hypothesized to be due
to stimulated echoes induced by pulse errors \cite{Franzoni2005,Franzoni2008,Franzoni2010,Franzoni2010a}
or due to the non-negligible effects of the interaction Hamiltonian
acting during the finite width pulses \cite{li_generating_2007,li_intrinsic_2008,dong_controlling_2008}.

The right hand panels show the same data, but plotted against the
number of pulses. They clearly show that the oscillation frequency
depends on the number of applied pulses or equivalently on the total
pulse-irradiation time. Similar oscillations have been also reported
in different samples \cite{li_generating_2007,li_intrinsic_2008,dong_controlling_2008,Franzoni2010}.
In our experiments, the oscillation pattern %
{} originates from the bimodal distribution of rf field amplitudes in
the coil. For our present analysis, the beating is not important because
it could be reversed \cite{dong_controlling_2008} or avoided by improving
the rf field coil. We instead concentrate on the decay of the envelope,
which represents the overall survival probability of the signal.

For the longitudinal initial state (upper panel), panel b shows that
the signal decay, as a function of the distance $\tau$ between successive
pulses, remains constant until $\tau=30\mu s$. The corresponding
cycle time is $\tau_{c}=2\tau+2\tau_{p}=80.8\mu s$, which is comparable
to the bath correlation time $\tau_{B}$; hence; the signal decay
for cases below $\tau=30\mu s$ is mainly due to pulse errors. For
longer delays $\tau_{c}>\tau_{B}$ ($\tau_{B}\sim110\mu$s), the decay
rate increases because of the reduction of time reversal efficiency
in the fluctuating environment. %
{} %

\subsection{CPMG-2}

\begin{figure}
\includegraphics[bb=10bp 10bp 250bp 315bp,clip]{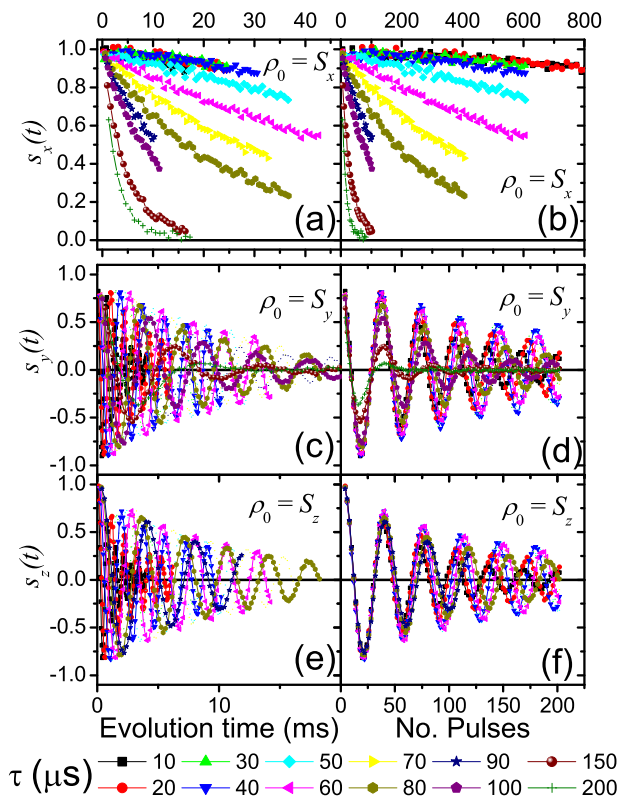}

\caption{(color online) Magnetization evolution for the CPMG-2 sequence. From
top to bottom the initial conditions are $\hat{S}_{x},$$\hat{S}_{y}$
and $\hat{S}_{z}$. The left-hand panels represent the $^{13}$C magnetization
as a function of the total evolution time, while right panels show
its evolution as a function of the number of applied pulses. The legend
at the bottom gives the delays $\tau$ between successive pulses.}

\label{Flo:CPMG2} 
\end{figure}

Figure \ref{Flo:CPMG2} shows the corresponding results for the CPMG-2
sequence. Since the first non-vanishing order of the Magnus expansion
for the CPMG-2 sequence (considering flip-angle errors) commutes with
$\hat{S}_{x}$, we expect that the signal decay for the $\hat{S}_{x}$
initial state is similar to that of the longitudinal initial state
of the CPMG experiments. The experimental results shown in (Fig. \ref{Flo:CPMG2})
clearly agree with this expectation. For the other initial states,
an oscillatory behavior similar to that for the transverse state of
the CPMG is observed. Although the oscillation still depends on the
number of pulses (right panels), the frequency is slower than in the
CPMG case, and the envelope of the oscillations decays more slowly.%
{}%
{} The origin of the oscillation pattern is again the bimodal distribution
of the inhomogeneity of the RF field generating an effective field
along the $x$ axis.%
{} The experimentally observed oscillation agrees with the results of
the effective nutation experiment. Again, we will concentrate on the
decay of the envelope. %

\subsection{PDD}

\begin{figure}
\includegraphics[bb=15bp 0bp 240bp 315bp,clip]{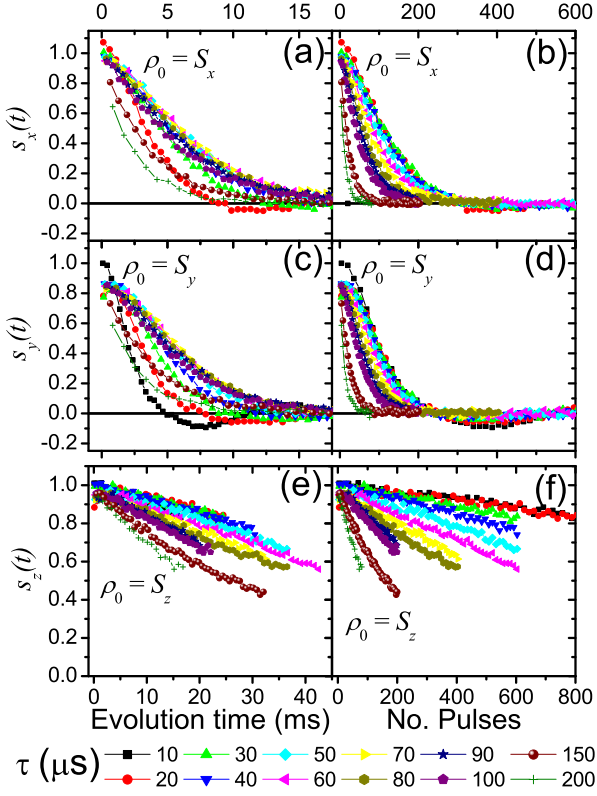}

\caption{(color online) Signal decay of the initial state of the qubit for
the PDD sequence. From top to bottom the initial conditions are $\hat{S}_{x},$$\hat{S}_{y}$
and $\hat{S}_{z}$. The left-hand panels represent the decay as a
function of the total evolution time while the right-hand panels show
the decay as a function of the number of applied pulses. The legend
at the bottom gives the delays $\tau$ between successive pulses.}

\label{Flo:PDD} 
\end{figure}

Figure \ref{Flo:PDD} shows the signal decay for different initial
conditions under the application of the PDD sequence of Fig. \ref{Flo:NMR_seq}(d).
One observes that the signal decay evolves qualitatively similar for
initial conditions in the plane transverse to the static field, i.e.
$\hat{S}_{x}$ and $\hat{S}_{y}$. This agrees with the theoretical
predictions: the sequence of evolutions $\hat{Y}f_{\tau}\hat{X}f_{\tau}\hat{Y}f_{\tau}\hat{X}f_{\tau}$
is nearly symmetric with respect to $x$ vs. $y$. The decays still
contain a small oscillatory contribution. Since it appears to depend
mostly on the number of pulses, rather than on the delays between
them, we attribute them to pulse errors that are not completely canceled.
Compared to CPMG and CPMG-2, the period of the oscillation is one
order of magnitude longer, indicating that the effect of the pulse
imperfections has been reduced by an order of magnitude. This general
improvement against pulse errors is because the sequence cancels the
zeroth order average Hamiltonian of the more general SE interaction
(\ref{eq:GSEI}), while the CPMG sequences cancel only its pure-dephasing
part.

In Fig. \ref{Flo:PDD}(b) and (d), the decay rates, in units of pulses,
up to $\tau=40\mu$s, i.e. $\tau_{c}=4\left(\tau+\tau_{p}\right)=201.6\mu$s,
are equal to within experimental error. This shows that the sequence
is more robust, compared to previous sequences, in the regime where
$\tau_{c}$ exceeds the bath-correlation time. However, from the time
evolution of the left panels a and c, the decay rate is larger than
the longitudinal case of CPMG or the $\hat{S_{x}}$ case of CPMG-2.

If the initial state is proportional to $\hat{S}_{z}$, its evolution
is qualitatively different. We believe that this results from the
fact that it is parallel to the static field and commutes with the
free precession Hamiltonian (\ref{eq:Hfreeevolution}). As a consequence,
this evolution reflects the implementation errors of the sequence.
The source of the decay is the pulse errors due to which the average
Hamiltonian no longer commutes with the initial state. This experiment
provides a means of quantifying pulse errors, and calibrating an optimal
setup of the sequence to enhance its performance.

\subsection{CDD}

The qualitative behavior of the CDD experiments is similar between
different orders and to the PDD one, but they change in the time scale
for which the initial state can be maintained. They are also more
robust against pulse errors -- the oscillation pattern is not observed.
For details of the survival probability evolution see the appendix
\ref{sec:CDD-experiments}. A summary of the results is presented
in Fig. \ref{Flo:Decay_times} in section \ref{sec:ComparisonsOptimal-choices}
where the decay times for different CDD orders and delays $\tau$
between pulses are plotted.

\subsection{UDD}

The experimental survival probabilities for the UDD sequences have
the same qualitative behavior as the CPMG curves. They manifest the
same asymmetries with respect to the initial state. That is expected
because UDD sequences also only reduce SE interactions of the form
(\ref{eq:HSE}). We observed that with increasing UDD order, the decay
rates increase, as predicted in Ref. \cite{UhrigLidar2010} for the
conditions satisfied in our experiments where $\left|b_{Sj}\right|\tau_{B}<1$
and $\tau_{c}\sim\tau_{B}$ . A summary of the rates is shown in the
next section in Fig. \ref{Flo:Decay_times}. An extensive analysis
of the performance of UDD sequences and non-equidistant pulse sequences
against equidistant ones for the present experimental conditions will
be given elsewhere \cite{UDDPaper}. %
{}

\section{Comparisons: optimal choices \label{sec:ComparisonsOptimal-choices}}

\begin{figure}
\includegraphics[bb=10bp 10bp 260bp 350bp,clip,scale=0.9]{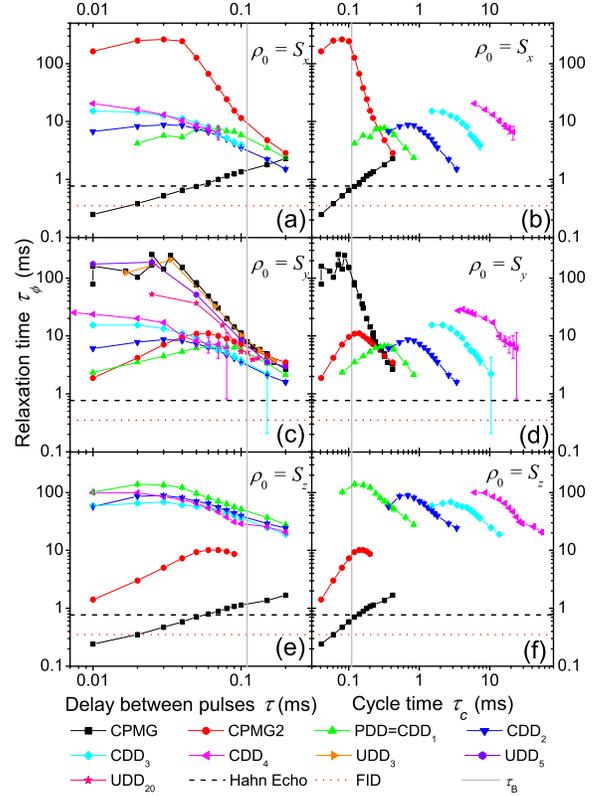}

\caption{(color online) Relaxation times of different initial conditions under
DD conditions as a function of the delay between pulses $\tau$ (left
panels) and the cycle time $\tau_{c}$ (right panels). From top to
bottom the initial condition is given by $\hat{\rho}_{0}=\hat{S}_{x}$,
$\hat{S}_{y}$ and $\hat{S}_{z}$ respectively. An optimal $\tau$
and consequently $\tau_{c}$ is observed for each sequence. The reduction
of the relaxation time to the right side of the optimal value is due
to the shifting environment: in this regime the cycle time is longer
than the correlation time of the bath, $\tau_{c}>\tau_{B}$. The reduction
for short cycle times indicates that in this regime, accumulated pulse
errors dominate.}

\label{Flo:Decay_times} 
\end{figure}

The goal of DD is the preservation of quantum states by the application
of suitable decoupling sequences. If the pulses are ideal and they
are applied with very short delays, it is possible to preserve quantum
states for arbitrarily long times in the presence of a system-environment
coupling that is linear in the system operators. However, for experiments
using non-ideal pulses, a finite cycle time optimizes the DD performance.
This can be seen very clearly in the summary of the experimental results
presented in Fig. \ref{Flo:Decay_times}. The left panels show the
DD decay times as a function of the delay $\tau$ for every sequence
and different initial conditions. As an example, for CPMG when the
initial condition is longitudinal to the pulses ($\hat{S}_{y}$),
the optimal cycle time is $\tau_{c}=80.8\mu s$ ($\tau=30\mu s$).
For longer cycle times, the decay time gets shorter, since the environment
changes during the cycle and the refocusing efficiency decreases.
\begin{figure}
\includegraphics[bb=15bp 10bp 270bp 240bp,clip,scale=0.8]{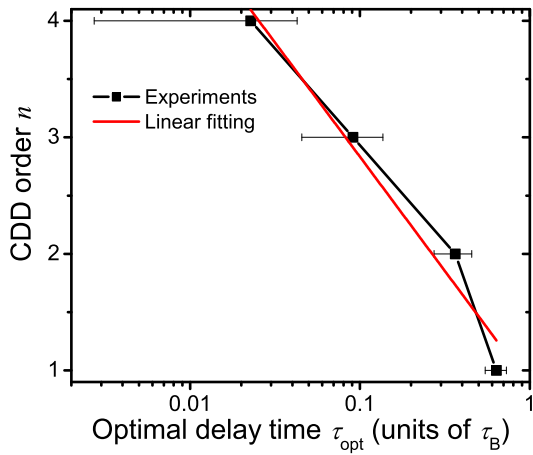}

\caption{(color online) CDD order $n$ as a function of its optimal delay between
pulses $\tau_{\mathrm{opt}}$ to reduce decoherence. The experimental
square points seem to satisfy a relation given by $n=c-b\ln(\tau_{\mathrm{opt}}/\tau_{B})$.
The solid (red) line shows a fitting curve with parameters $c=(0.9\pm0.2)$
and $b=(-0.9\pm0.1)$.}

\label{Flo:cddordervstau_opt} 
\end{figure}

While shorter cycle times should give even better results under ideal
conditions, we find experimentally a decrease of the relaxation time.
This can be attributed to an accumulation of pulse errors, which dominates
in this regime. Similar results are observed for the CPMG-2 if we
exchange $\hat{S}_{y}$ with $\hat{S}_{x}$. This can be seen clearly
under conditions of transverse initial states ($\hat{S}_{x}$ for
CPMG) where the decay time is proportional to the cycle time. This
means that the error per cycle is independent of the cycle time and
corresponds thus to a zero-order term of the average Hamiltonian.
This is the behavior expected for flip angle errors, which are the
main source of the decay in this regime. Since flip angle errors are
in no way compensated for transverse states in the CPMG sequence,
their accumulated effect is so strong that the optimal cycle time
exceeds $\tau_{B}$ and the sequence performs only marginally better
than the Hahn echo sequence, which has the longest cycle time.%
{}

If we consider the CPMG-2 sequence with the initial conditions $\hat{S}_{y}$
and $\hat{S}_{z}$, the decay time grows $\propto\tau_{c}^{2}$ for
short times. This implies that in this case, the dominant error term
is proportional to $\tau_{c}$, i.e. it corresponds to a first-order
term of the average Hamiltonian. Moreover, its optimal relaxation
time is one order of magnitude longer than the Hahn echo decay time.

The behavior of the UDD sequences is similar to that of CPMG. We show
here only their decay times for $\hat{S}_{y}$ as initial condition
{[}Fig. \ref{Flo:Decay_times}(c){]}, i.e. longitudinal to the DD
pulses. They are plotted as a function of the average delay between
pulses. The figure shows that the UDD decay times are always shorter
than those of CPMG. Moreover, increasing the UDD order reduces the
decay time, as expected by theoretical expectations when $\left|b_{Sj}\right|\tau_{B}<1$
and $\tau_{c}\sim\tau_{B}$ \cite{UhrigLidar2010}. A regime where
UDD performs better than CPMG may perhaps exist at short cycle times
compared with $\tau_{B}$ \cite{UDDPaper}, provided the pulse errors
can be made sufficiently small that they do not dominate over external
sources of decoherence. Recent proposals of UDD based sequences that
reduce decoherence of a general SE interaction like Eq. (\ref{eq:GSEI}),
could allow one to find this regime \cite{Uhrig2009,west_near-optimal_2010}.

For PDD the optimal cycle time for $\hat{S}_{x}$ and $\hat{S}_{y}$
is $\tau_{c}=321.6\mu$s, ($\tau=70\mu$s), which is longer than $\tau_{B}$.
The resulting performance is relatively poor. The resulting decay
time is similar to that of CPMG for the same cycle time {[}Fig. \ref{Flo:Decay_times}(d){]}
or CPMG-2 for the $\hat{S}_{x}$ initial condition {[}Fig. \ref{Flo:Decay_times}(b){]}.
However, for these particular initial conditions CPMG and CPMG-2 can
be made to perform an order of magnitude better by reducing the cycle
time.

For CDD$_{2}$ with initial conditions transverse to the static field,
the optimal cycle time is $\tau_{c}=16\tau+20\tau_{p}=688\mu\mbox{s}$,
i.e. $\tau=30\mu$s. For CDD$_{3}$ and CDD$_{4}$ the shortest delay
time between pulses of $\tau=10\mu$s and $\tau=2.5\mu$s are the
optimal situations, giving $\tau_{c}(\mbox{CDD}_{3})=64\tau+84\tau_{p}=1513.6\mu$s
and $\tau_{c}(\mbox{CDD}_{4})=256\tau+388\tau_{p}=4675.2\mu$s. The
optimal delay $\tau$ becomes shorter with increasing CDD order, because
the cycle time increases by a factor of 4 for each level of concatenation.
Apparently, the pulse errors do not accumulate as strongly as in the
case of CPMG, which may be attributed to the fact that CDD is designed
to compensate pulse errors \cite{khodjasteh_fault-tolerant_2005,khodjasteh_performance_2007}.
The crossover cycle time, where the transition occurs from a decay
dominated by pulse errors to the regime where the decay is dominated
by the short bath correlation time is increased, as shown in the right
hand panels. Even for cycle times that are much longer than the bath-correlation
time, the CDD provides a significant reduction of the decoherence
rate compared to the free evolution decay (dotted lines) and the Hahn
echo decay (dashed lines).%
{} An optimal cycle time when considering imperfect finite pulses was
predicted by Khodjasteh and Lidar \cite{khodjasteh_performance_2007}.
Figure \ref{Flo:cddordervstau_opt} shows the experimental relation
between the optimal delays $\tau$ and their respective CDD order
$n$ (square points). It seems to satisfy a relation given by $n=c-b\ln(\tau_{\mathrm{opt}}/\tau_{B})$,
where $c$ and $b$ are constants and $\tau_{\mathrm{opt}}$ is the
optimal delay for a given $n$ (see below).

The unifying result of the curves shown in the left-hand panels of
Figure \ref{Flo:Decay_times} is that the optimal delay between pulses
is always shorter than the bath correlation time, with comparable
values for all sequences, with the single exception of the CPMG sequence
for initial conditions $\hat{S}_{x}$ and $\hat{S}_{z}$, as discussed
above. Clearly, this timescale is determined by the (average) delay
between pulses $\tau$, not by the cycle time $\tau_{c}$. Expressing
this differently, one might say that only a small fraction of what
is lost in a single echo can be refocused by compensated sequences.
If we look at pulse spacings longer than the bath correlation time
$\tau_{B}$, the differences between sequences become very small and
the decay times approximate those of FID and Hahn echo. Accordingly,
it appears important to keep the number of pulses per unit time constant
when comparing different DD sequences.

\begin{figure}[b]
 \includegraphics[bb=10bp 10bp 290bp 350bp,clip,scale=0.8]{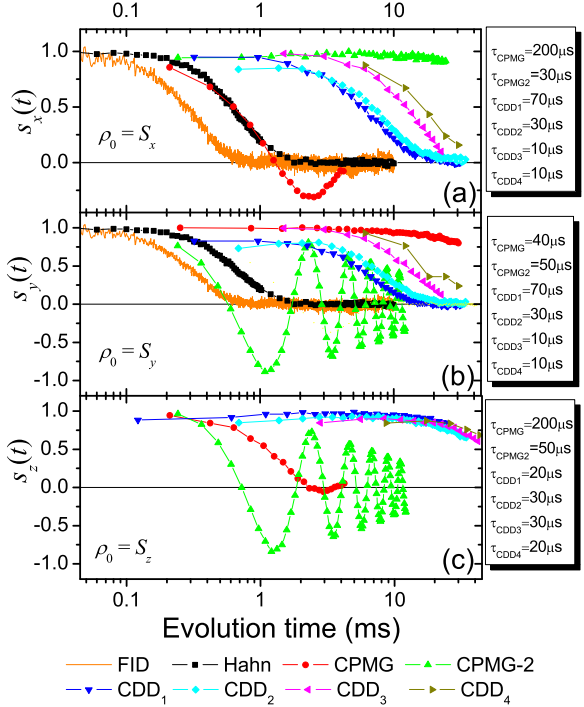}

\caption{(color online) Time evolution of the survival probability for the
optimal cycle times of the different DD sequences. From top to bottom
the initial states are $\hat{S}_{x},$$\hat{S}_{y}$ and $\hat{S}_{z}$.
The optimal delays $\tau$ are given in the legends. }

\label{Flo:Comparisons} 
\end{figure}

We now compare the different DD sequences with the optimal cycle time
for each sequence. Figure \ref{Flo:Comparisons} shows the evolution
of the survival probabilities for different initial conditions for
all the sequences discussed here.

As a general rule, we note that for increasing CDD order, the optimal
pulse delay $\tau$ gets shorter and for longer delays between pulses,
higher CDD orders do not perform better than lower CDD orders. Hence,
keeping the delay between pulses constant, there is an optimal CDD
order for reducing decoherence as predicted in Refs. \cite{khodjasteh_performance_2007,Ng2009}.
It is difficult to find accurately the optimal CDD order as a function
of $\tau$ from Figs. \ref{Flo:Decay_times} (a) and (c) in order
to compare with the theoretical predictions of Eq. (140) in Ref. \cite{Ng2009}.
However, the relation given in Fig. \ref{Flo:cddordervstau_opt} for
the optimal delay between pulses $\tau_{\mathrm{opt}}$ behaves similar.
A linear fitting of the experimental data gives $n=c-b\ln(\tau_{\mathrm{opt}}/\tau_{B})$
with $c=(0.9\pm0.2)$ and $b=(-0.9\pm0.1)$ agreeing well with the
predicted expression \cite{Ng2009}.

The best DD sequence and its corresponding optimal cycle time depends
on hardware limitations, and importantly on the desired goal. If one
aims to freeze a quantum state during a short time, its value will
bound the cycle time and as a result, the maximal CDD order that can
be applied. For longer times, increasing the CDD order will be advantageous,
but power dissipation may force a reduction in the number of pulses
and simultaneously the CDD order. For specific initial conditions,
CPMG and CPMG-2 are the best choices for reducing decoherence; however
the large asymmetry of these sequences to other initial conditions
limits their usefulness when the initial state of the qubit is unknown.
Note that in these cases, it has been shown that coherences could
be frozen as labelled polarization \cite{Franzoni2010a}. If the goal
is the preservation of an unknown quantum state, the CDD sequences
provide the best overall performance.

While the SE interaction produces pure dephasing to the spin, in principle
DD sequences that compensate pure dephasing decoherence should be
sufficient. In consequence concatenating sequences like $f_{\tau}\hat{Y}f_{\tau}\hat{Y}$,
which reduce pure dephasing processes, could be beneficial. However,
the finite precision of control pulses generates an effective Hamiltonian
of the form (\ref{eq:GSEI}), and thus sequences developed to reduce
pure dephasing processes have asymmetric performances against initial
states directions, i.e. they do not generate a unit evolution operator
of the qubit. Thus, we compared, as a test bed, CPMG/UDD sequence
that compensate pure dephasing with XY-4 \cite{Maudsley1986,Gullion1990}
based sequences that compensate a general interaction in order to
show their effects against pulse errors.

From our results it is evident that for short delays the main source
of DD decays are static pulse errors; this limits the maximal performance.
However, CPMG and CPMG-2 show the potentially achievable DD performance
if the pulse errors are reduced. This implies that new DD proposals
should focus on the compensation of pulse errors for these kinds of
experimental conditions, similar to the proposal by Viola and Knill
\cite{Viola2003} or Uhrig and Pasini \cite{Uhrig2010,Pasini2009}.
CDD-type sequences do compensate for pulse errors, but only at the
end of the CDD cycle. This limits their performance because of the
exponential growth of the cycle time with the CDD order. As an alternative
method, we suggest to find sequences that compensate pulse errors
to zero order during each step of the concatenation procedure. That
would be advantageous because the zero order compensation cycle time
remains constant and equal to the PDD cycle time, as assumed for ideal
pulses.%
{}

\section{Conclusions}

We have experimentally applied different dynamical decoupling sequences
to a qubit-system coupled to a spin-bath in order to test and compare
their performance. The system used is typical for spin-based solid-state
systems where the spectral density of the bath is given by a normal
(Gaussian) distribution and the system-environment interaction is
weak compared with the intra-bath interaction. The experiments were
performed in the regime where the average spacing between the pulses
is comparable to the bath-correlation time. This article focuses on
measuring and fighting decoherence, and the results do not depend
on the readout or initialization scheme used for that purpose. Thus,
the results should apply directly to other spin-based quantum information
processing systems, such as electron spins in diamonds \cite{Naydenov2010,Lange2010,Ryan2010},
electron spins in quantum dots \cite{Hanson2007,Bluhm2010,Barthel2010}
and donors in silicon \cite{Kane1998,Morton2008}.

While the design of DD sequences is typically based on the assumption
that the cycle times is shorter than the bath-correlation time, we
demonstrated that even without satisfying this condition, dynamical
decoupling reduces decoherence significantly. We showed that the main
limitation to the reduction of the DD decay rates is due to the finite
precision of the control operations -- in our system, flip-angle errors
were the main source. Therefore, CPMG or UDD-type sequences that reduce
purely-dephasing or purely-dissipative interactions with the bath
perform well only for specific initial conditions of the qubit ensemble.
For the privileged initial condition, CPMG-type sequences performed
better than any other DD sequence. But, if the goal is to approach
a unit evolution operator, PDD sequence and its concatenated form
(CDD) are the best overall option. In agreement with previous predictions
\cite{khodjasteh_performance_2007,Ng2009,Hodgson2010} our results
show that, depending on limitations of hardware and the required time
to keep the initial state coherent, increasing the CDD order is not
always useful. There is an optimal CDD order depending on the power
available for the control pulses and their finite precision. We present
strong evidence that in order to improve dynamical decoupling sequences,
they should be designed to compensate pulse errors. 
\begin{acknowledgments}
This work is supported by the DFG through Su 192/24-1. GAA thanks
the Alexander von Humboldt Foundation for a Research Scientist Fellowship.
We thank Daniel Lidar for helpful discussions and Marko Lovric and
Ingo Niemeyer for technical support. 
\end{acknowledgments}
\appendix

\section{CDD experiments\label{sec:CDD-experiments}}

\begin{figure}
\includegraphics[bb=15bp 10bp 240bp 315bp,clip]{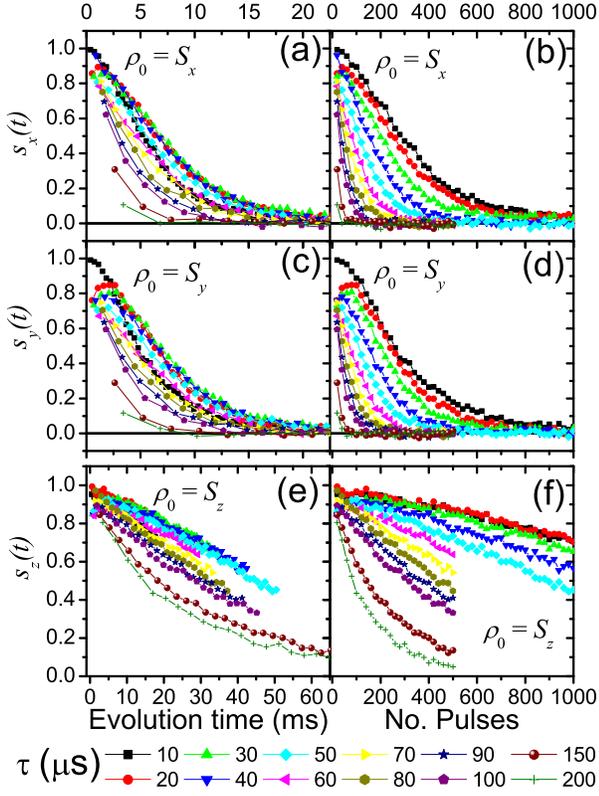}

\caption{(color online) Signal decay of the initial state of the qubit for
the CDD$_{2}$ sequence. From top to bottom the initial conditions
are $\hat{S}_{x},$$\hat{S}_{y}$ and $\hat{S}_{z}$. The left-hand
panels represent the decay as a function of the total evolution time
while the right-hand panels show the decay as a function of the number
of applied pulses. The legend at the bottom gives the delays $\tau$
between successive pulses.}

\label{Flo:CDD2} 
\end{figure}

\begin{figure}
\includegraphics[bb=15bp 10bp 240bp 315bp,clip]{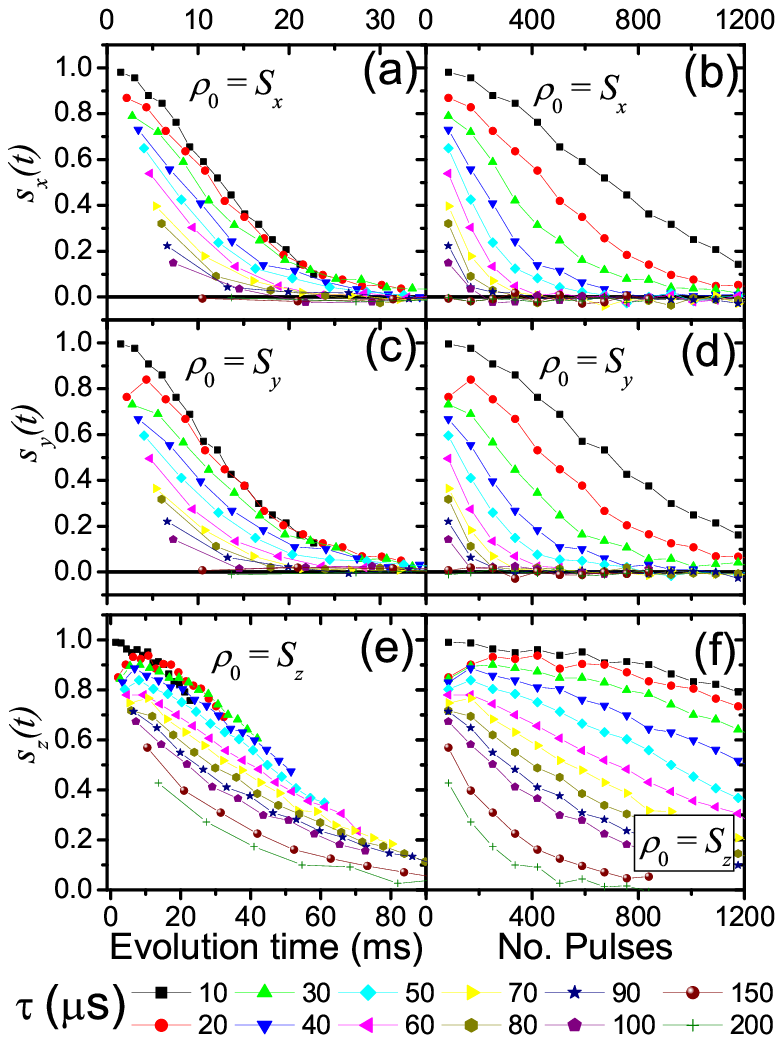}

\caption{(color online) Signal decay of the initial state of the qubit for
the CDD$_{3}$ sequence. From top to bottom the initial conditions
are $\hat{S}_{x},$$\hat{S}_{y}$ and $\hat{S}_{z}$. The left-hand
panels represent the decay as a function of the total evolution time
while right panels the decay as a function of the number of applied
pulses. The legend at the bottom gives the delays $\tau$ between
successive pulses.}

\label{Flo:CDD3} 
\end{figure}

\begin{figure}
\includegraphics[bb=15bp 10bp 240bp 315bp,clip]{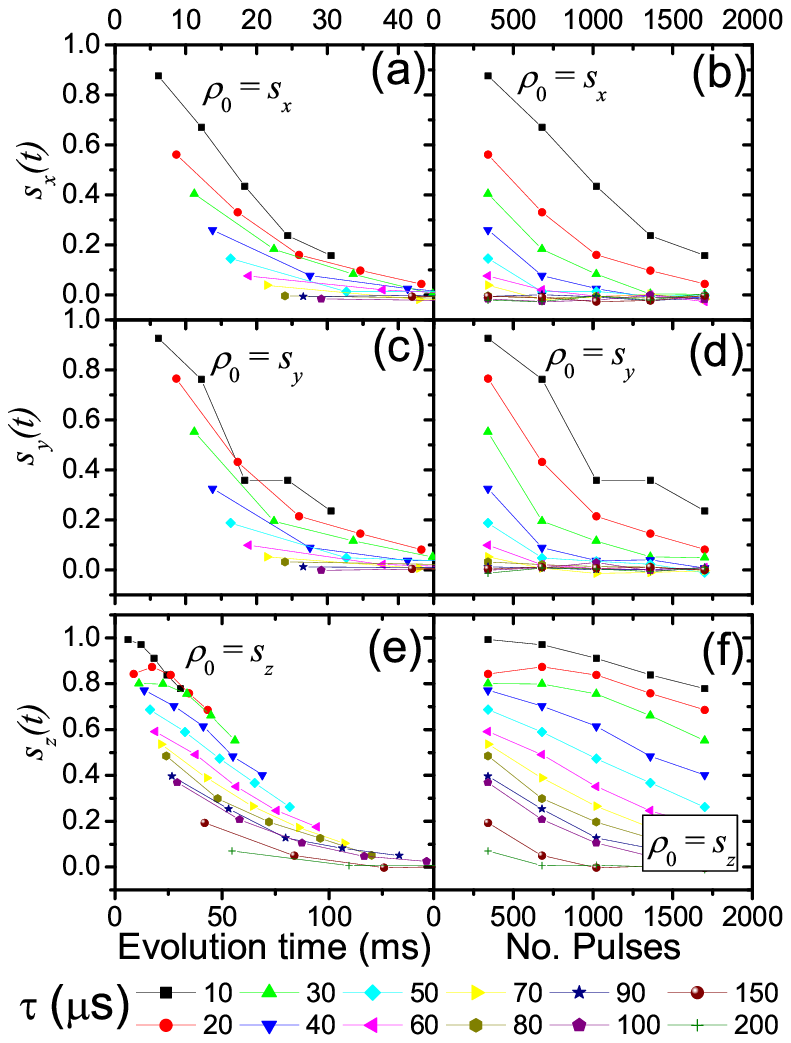}

\caption{(color online) Signal decay of the initial state of the qubit for
the CDD$_{4}$ sequence. From top to bottom the initial conditions
are $\hat{S}_{x},$$\hat{S}_{y}$ and $\hat{S}_{z}$. The left-hand
panels represent the decay as a function of the total evolution time
while the right-hand panels show the decay as a function of the number
of applied pulses. The legend at the bottom gives the delays $\tau$
between successive pulses.}

\label{Flo:CDD4} 
\end{figure}

Figures \ref{Flo:CDD2}, \ref{Flo:CDD3} and \ref{Flo:CDD4} show
the experimentally observed signal decays for CDD$_{2}$, CDD$_{3}$
and CDD$_{4}$ respectively.%

\bibliographystyle{apsrev} \bibliographystyle{apsrev}\bibliographystyle{apsrev}
\bibliographystyle{apsrev}
\bibliography{CDD,bibliography,CDD2}

\end{document}